\documentclass[aps,pre,10pt,twocolumn,superscriptaddress,showpacs,floatfix]{revtex4-1}
\usepackage{graphicx}
\usepackage{dcolumn}
\usepackage{bm}
\usepackage{amsmath}

\begin{document}

\title{Radiation influence on the temperature-dependent parameters of fluids}

\author{L. A. Bulavin}
 \affiliation{Physics Faculty, Taras Shevchenko National University of Kyiv, 4 Glushkova Av.,  Kyiv, 03022, Ukraine}
\author{K. V. Cherevko}
 \affiliation{Physics Faculty, Taras Shevchenko National University of Kyiv, 4 Glushkova Av.,  Kyiv, 03022, Ukraine}
\author{D. A. Gavryushenko}
 \affiliation{Physics Faculty, Taras Shevchenko National University of Kyiv, 4 Glushkova Av.,  Kyiv, 03022, Ukraine}
\author{V. M. Sysoev}
 \affiliation{Physics Faculty, Taras Shevchenko National University of Kyiv, 4 Glushkova Av., Kyiv, 03022, Ukraine}
\author{T. S. Vlasenko}
 \email{vlasenko.tata@gmail.com}
 \affiliation{Institute for Safety Problems of Nuclear Power Plants, National Academy of Sciences of Ukraine, 12 Lysogirska St., Kyiv, 03028, Ukraine}

\date{\today}

\begin{abstract}
Based on the fundamental Bogolyubov chain of equations, a model relating the  structural and thermophysical properties of the nonequilibrium liquid systems under irradiation in stationary state is introduced. The obtained results suggest that the thermophysical properties of the liquid systems under irradiation are defined by the {\textquotedblleft}effective temperature{\textquotedblright} that can be calculated from the perturbed momentum distribution functions of the systems. It is shown that the structural changes in the liquid systems under irradiation are caused by the changes in the coefficients of the Maxwell distribution function due to the momentum exchange between the active particles and the particles forming the liquid. To confirm the theoretical predictions, a qualitative comparison of the model with the existing experimental data on irradiation influence on the surface tension coefficients of liquids is performed.
\end{abstract}\texttt{}


\maketitle

\section{Introduction.}
Over the past few decades numerous studies of the influence of irradiation on different physical systems have appeared in the literature \cite{Montenegro2005, Parajon2008, Draganic2005, Kreipl2009, Palfi2010}. In most of the papers devoted to the physical mechanisms involved in the interaction the solid state  is concerned \cite{Urusov2012, Zarkadoula2013, Malerba2010, Trachenko2012}, but only few of them are devoted to studies of the physical mechanisms the irradiation affects the liquid matter. This is surprising since understanding the physical mechanisms of the interaction of irradiation with the liquid matter is important both from theoretical point of view and for various applications, namely in medicine, nuclear engineering (molten salt reactors), etc. Contemporary models of irradiation influence on the soft matter \cite{Wai1998} suggest that the absorbed radiation energy ionizes atoms and molecules. Thus, irradiation leads to the destruction of the bonds and formation of the unstable active ions and radicals. Consequently, those radiolysis products define the redox potential of the system and form unstable substructures (tracks), changing the instant structure of the liquid \cite{Fisher1961}. When the system is exposed to irradiation during long enough time, the concentration of radiolysis products reaches some stationary level and the system comes to a nonequilibrium stationary state. As an example of such a system one can mention the water solution of $RaCl_2$. In that case dissociated $Ra$ ions irradiate $\alpha-$, $\beta-$ and $\gamma$ particles, bringing the system into a nonequilibrium stationary state. Another example is a chemical nuclear reaction in the homogenous system with uniformly distributed active particles.

It is evident that such structural changes should cause changes of the thermodynamic properties of the liquid system. Among theses one can mention both the properties related to the permanent structure of the liquid (such as surface tension or viscosity) and to the instant structure (such as electroconductivity) \cite{Martino2006, Zenkiewicz2007, Byung2008, Weon2008}. Within the formalism of the nonequilibrium thermodynamics, such changes may be explained by entropy production in relaxation processes (e.g., structural relaxation) occurring in the system to compensate the negative entropy introduced by active particles \cite{Zubarev1974, Groot1984, Perez-Madrid2007, Ruelle2003, Gallavotti2004}. Therefore, it might be interesting to study the structural changes in the liquid matter under irradiation, as well as the origin of the relaxation processes and their influence on the dynamic values of the thermodynamic parameters.

The complexity of the experiments that may provide reliable information regarding structural changes and changes of the corresponding thermodynamic properties of the liquid systems under irradiation should be also mentioned. It results in a situation where, in spite of the large number of papers describing the consequences of the liquids exposure to radiation \cite{Kolesnichenko1975} and the existing chemical models of the process (e.g., radiolysis \cite{Draganic2005, Burns1989, Sims2006}), the physical nature of the behavior of the liquid under irradiation is still uncertain. It should be also mentioned that we still lack a general model to quantify the radiation effects in liquids.

The aim of this work is to use the fundamental laws of the nonequilibrium statistical thermodynamics in the studies of the physical mechanisms responsible for the changes in the structural and thermodynamic properties of the liquid systems under irradiation.

\section{Theoretical model.}

Here we treat the influence of irradiation on the liquid system as momentum transfer between active particles and those forming the liquid. Therefore, we assume that the main mechanism causing the changes in the structural and thermodynamic properties of the liquid systems under irradiation is the deviation of the velocity distribution function from the Maxwell distribution, typical of the equilibrium state \cite{Kolesnichenko1975}. Hence, the coefficients $A$ and $\phi$ in the exponential Maxwell distribution
\begin{equation}\label{Maxw}
f(p)=A{\exp(-\phi{p}^2)}
\end{equation}
under the irradiation should differ from those obtained from the equation of state in the equilibrium state.

To justify the suggested approach, we note that there exists a number of papers devoted to  studies of the properties of the nonequilibrium systems based on the nonequilibrium potential method, the dynamic pair correlation function \cite{Kalyuzhnyi1999, Gan1991} or different expansions of the momentum distribution functions about the local Maxwellian distributions \cite{Loose1988}. Among them are studies of the behavior of a gas undergoing a plane Couette flow. In that case the well known expansion functions are used to modify the Maxwell distribution that are Sonine polynomials. Another example is the study of the chemically reacting gas that require corrections to the Maxwellian distribution \cite{Takayanagi1951}. Such a system has much in common with the liquid system under irradiation as in the radiolysis process in liquids there appear a large number of interacting particles \cite{Draganic2005}. Deviations from the standard Maxwellian distribution are also intensively studied in the nuclear astrophysics community as they are important in the description of the nuclear reactions that occur in stars \cite{He2013,DeglInnocentia1998}. When discussing deviations from the Maxwellian distribution in the stationary states one can also mention the power-law velocity distribution in granular gases \cite{Bennaim2005, Bennaim2005a} as well as the non-Maxwell velocity distributions observed in liquids in the stationary state in an external field \cite{Alastuey2010, Gervois1986,Piasecki1979,Zhu1989,Zhu1990}. Some of the above problems seem to have much in common with the problem of the liquid systems exposure to irradiation. Therefore, the idea to study the influence of the changes in the velocity distribution function of the liquid systems under irradiation on their structural and thermodynamic properties seems to be attractive.

To describe the structural changes in the liquid system under irradiation we suggest to use the fundamental Bogolyubov chain of equations \cite{Bogolyubov1962,Gurov1966}:

\begin{widetext}
\begin{multline}
\label{Bogolyubov1}
\frac{\partial F_{n}({{\bf r}_1},{{\bf r}_2},...,{{\bf r}_n},{{\bf p}_1},{{\bf p}_2},...,{{\bf p}_n},t)}{\partial t}
=[H^{(n)},F_n({{\bf r}_1},{{\bf r}_2},...,{{\bf r}_n},{{\bf p}_1},{{\bf p}_2},...,{{\bf p}_n},t)]\\
+\rho\sum_{i=1}^{n}\int\frac{\partial \Phi(|{{\bf r}_i}-{{\bf r}_{n+1}}|)}{\partial {\bf r}_i}
\times\frac{\partial F_{n+1}({{\bf r}_1},{{\bf r}_2},...,{{\bf r}_{n+1}},{{\bf p}_1},{{\bf p}_2},...,{{\bf p}_{n+1}},t)}{\partial {\bf p}_i}d{\bf r}_{n+1}d{\bf p}_{n+1},
\end{multline}
\end{widetext}
where $H^{(n)}$ is the Hamiltonian of the closed $n$-particle system; $F_{n}({{\bf r}_1},{{\bf r}_2},...,{{\bf r}_n},{{\bf p}_1},{{\bf p}_2},...,{{\bf p}_n},t)$ \t is the nonequilibrium $n$-th order distribution function depending on space coordinates ${{\bf r}_1},{{\bf r}_2},...,{{\bf r}_n}$,  momentum ${{\bf p}_1},{{\bf p}_2},...,{{\bf p}_n}$ and time $t$;  $\Phi(|{{\bf r}_i}-{{\bf r}_{n+1}}|)$ \t is the potential of interaction between $i$-th and $(n+1)$-th particles; $\rho=\frac {N} {V}$ \t is the numerical density.

 From Eq. (\ref{Bogolyubov1}), under the condition of thermodynamic equilibrium a well known equation for the equilibrium pair distribution function $F_{2}({\bf r},{\bf p},t)$ can be easily obtained, that, in turn, allows us to calculate the equilibrium thermodynamic properties of the system.

At this point, it should be noted that the interaction of irradiation and high energy particles, in particular with the liquid matter, involves abrupt changes in the particles velocity. It means that the phase volume is not conserved, making impossible the use of Hamiltonian mechanics to describe the process and, therefore, the use of the Bogolyubov chain of equations is unjustified. In the present work we do not dwell in the extremely complicated evolution of the system from the strongly nonequilibrium state (right after the beginning of the irradiation) toward the stationary nonequilibrium, when the system passes a number of different intermediate states characterized by different sets of relaxation times. In the general case, after perturbation, the system moves toward an equilibrium state. In our case, constant irradiation does not allow the system to evolve to this state but rather moves it to some stationary nonequilibrium state. Following the ideas of Bogolyubov \cite{Kac1959}, we assume that, after some period of time, sufficient for a few  collisions, chaotization of the particles movement is observed. Thereupon, a kind of equilibrium can be observed in the velocity distribution, and the evolution of the system is governed by  changes of the macroscopic parameters  and external factors. Therefore, our model is applicable for the systems when the state of the system is defined not by the path used to reach it but rather by the macroscopic parameters and the principal of the minimum entropy production with fixed external parameters that do not allow the system to come to equilibrium \cite{Groot1984, Prigogine1947, Zubarev1997}. Then it is legitimate to use Eq. \ref{Bogolyubov1} with $F_{n}({{\bf r}_1},{{\bf r}_2},...,{{\bf r}_n},{{\bf p}_1},{{\bf p}_2},...,{{\bf p}_n},t)=F_{n}({{\bf r}_1},{{\bf r}_2},...,{{\bf r}_n},{{\bf p}_1},{{\bf p}_2},...,{{\bf p}_n},\rho(t),T(t),Ext(t)),$ where the dependence on time is included in the macroscopic parameters such as density $\rho(t)$, temperature $T(t)$, and external influence $Ext(t)$.

For a homogeneous system in the stationary state, a function $F_{2}({\bf r},{\bf p},t)$ can be taken as the factor \cite{Fisher1961} of the spacial part $g_2({\bf{r}})$ and the momentum part $f_2({\bf{p}})$ ($F_{2}({\bf r},{\bf p},t)=f_2({\bf{p}})g_2({\bf{r}})$). In the equilibrium case, when $f_{2}({\bf p})$ is the Maxwellian function, the problem becomes trivial. Therefore, it is interesting to study the case of a non-Maxwellian $f_{2}({\bf p}),$ typical of the nonequilibrium stationary systems, that can be found in the liquid systems under irradiation.

To describe the structure and the thermophysical properties of the system, it is sufficient to know the time-independent pair distribution function $g_{2}({{\bf r}_1},{{\bf r}_2})$ with ${{\bf r}_1},{{\bf r}_2}$ giving the positions of the centers of the molecules. Therefore, for the vast majority of the applications, one can cut the chain of Eq. (\ref{Bogolyubov1}) at the second equation.

There exist quite a number of approaches to find the pair distribution function. Among them are  experimental, theoretical, and computer simulation methods. At the same time, most of them are applicable only in the equilibrium case and, therefore, are not suitable for the case of the liquid matter under irradiation when the system is in the nonequilibrium stationary state. Our objective here is calculation of the pair distribution function $g_{2}({{\bf r}_1},{{\bf r}_2})$ for a nonequilibrium stationary state. According to our assumption that state is characterized by the deformed distribution function in the momentum space due to irradiation.

\subsection{Single-component system}
In order to develop a clear physical picture of the phenomena, in this paper a detailed analysis for the single-component liquid system under irradiation is presented. For the single-component system in the stationary nonequilibrium state characterized by the minimal entropy production with fixed macroscopic parameters ($T=const$, $\rho=const$) and constant irradiation rate $Ext=const$ one obtains $\frac{\partial F_{n}({{\bf r}_1},{{\bf r}_2},{{\bf p}_1},{{\bf p}_2},\rho,T,Ext)}{\partial t}=0$. Then, the second equation of the Bogolyubov chain (\ref{Bogolyubov1}) reads:
\begin{widetext}
\begin{multline}
\label{Bogolyubov2}
\frac{{\bf p}_1}{m}\frac{\partial {F_2}({{\bf r}_1},{{\bf r}_2},{{\bf p}_1},{{\bf p}_2})}{\partial {\bf r}_1})+\frac{{\bf p}_2}{m}\frac{\partial {F_2}({{\bf r}_1},{{\bf r}_2},{{\bf p}_1},{{\bf p}_2})}{\partial {\bf r}_2}
-\frac{\partial \Phi(|{{\bf r}_1}-{{\bf r}_2}|)}{\partial {\bf r}_1}\frac{\partial {F_2}({{\bf r}_1},{{\bf r}_2},{{\bf p}_1},{{\bf p}_2})}{\partial {\bf p}_1}
-\frac{\partial \Phi(|{{\bf r}_1}-{{\bf r}_2}|)}{\partial {\bf r}_2}\frac{\partial {F_2}({{\bf r}_1},{{\bf r}_2},{{\bf p}_1},{{\bf p}_2})}{\partial {\bf p}_2}\\
=\rho \int\left(\frac{\partial\Phi(|{{\bf r}_1}-{{\bf r}_3}|)}{\partial {\bf r}_1}
\frac{\partial {F_3}({{\bf r}_1},{{\bf r}_2},{{\bf r}_3},{{\bf p}_1},{{\bf p}_2},{{\bf p}_3})}{\partial {\bf p}_1}
+\frac{\partial \Phi(|{{\bf r}_2}-{{\bf r}_3}|)}{\partial {\bf r}_2}
\frac{\partial {F_3}({{\bf r}_1},{{\bf r}_2},{{\bf r}_3},{{\bf p}_1},{{\bf p}_2},{{\bf p}_3})}{\partial {\bf p}_2}\right)d{\bf r}_3d{\bf p}_3,
\end{multline}
Separating the variables in Eq. (\ref{Bogolyubov2}) and accounting for $F_2({\bf{r}},{\bf{p}})=g_2({\bf{r}})f_2({\bf{p}})$, it is possible to write two equations:
\begin{multline}
\label{Bogolyubov3}
\left(\frac{\partial \Phi(|{{\bf r}_1}-{{\bf r}_2}|)}{\partial {\bf r}_1}\frac{\partial {f_2}({{\bf p}_1},{{\bf p}_2})}{\partial {\bf p}_1}{g_2}({{\bf r}_1},{{\bf r}_2})-\frac{{\bf p}_1}{m}{\frac{\partial g_{2}({{\bf r}_1},{{\bf r}_2})}{\partial {\bf r}_1}}{f_2}({{\bf p}_1},{{\bf p}_2})\right)
+\rho \int\frac{\partial \Phi(|{{\bf r}_1}-{{\bf r}_3}|)}{\partial {\bf r}_1}\frac{\partial {f_3}({{\bf p}_1},{{\bf p}_2},{{\bf p}_3})}{\partial {\bf p}_1}\\
\times g_{3}({{\bf r}_1},{{\bf r}_2},{{\bf r}_3}){d{\bf r}_3}{d{\bf p}_3}=0,
\end{multline}

\begin{multline}
\label{Bogolyubov3a}
\left(\frac{\partial \Phi(|{{\bf r}_1}-{{\bf r}_2}|)}{\partial {\bf r}_2}\frac{\partial {f_2}({{\bf p}_1},{{\bf p}_2})}{\partial {\bf p}_2}{g_2}({{\bf r}_1},{{\bf r}_2})-\frac{{\bf p}_2}{m}{\frac{\partial g_{2}({{\bf r}_1},{{\bf r}_2})}{\partial {\bf r}_2}}{f_2}({{\bf p}_1},{{\bf p}_2})\right)
+\rho \int\frac{\partial \Phi(|{{\bf r}_2}-{{\bf r}_3}|)}{\partial {\bf r}_2}\frac{\partial {f_3}({{\bf p}_1},{{\bf p}_2},{{\bf p}_3})}{\partial {\bf p}_2}\\
\times g_{3}({{\bf r}_1},{{\bf r}_2},{{\bf r}_3}){d{\bf r}_3}{d{\bf p}_3}=0.
\end{multline}
\end{widetext}
Integration of one of Eqs. (\ref{Bogolyubov3}) and (\ref{Bogolyubov3a}) with respect to ${\bf{p_1}}$ and ${\bf{p_2}}$ and accounting for $\int d{{\bf{p}}_3}f_3 ({{\bf{p}}_1},{{\bf{p}}_2},{{\bf{p}}_3})=f_2 ({{\bf{p}}_1},{{\bf{p}}_2})$ gives the modified Bogoliubov {\textendash} Born {\textendash} Green {\textendash} Kirkwood {\textendash} Yvon (BBGKY) equation for the nonequilibrium stationary case
\begin{widetext}
\begin{multline}
\label{Bogolyubov4}
-\frac{\partial g_{2}({{\bf r}_1},{{\bf r}_2})}{\partial {\bf r}_1}\int\frac{{\bf p}_1}{m}{f_2}({{\bf p}_1},{{\bf p}_2})\partial{{\bf p}_1}\partial{{\bf p}_2}+\frac{\partial \Phi(|{{\bf r}_1}-{{\bf r}_2}|)}{\partial {\bf r}_1}g_{2}({{\bf r}_1},{{\bf r}_2})\int\frac{\partial {f_2}({{\bf p}_1},{{\bf p}_2})}{\partial {\bf p}_1}\partial{{\bf p}_1}\partial{{\bf p}_2}\\
+\rho \int\frac{\partial \Phi(|{{\bf r}_1}-{{\bf r}_3}|)}{\partial {\bf r}_1}g_{3}({{\bf r}_1},{{\bf r}_2},{{\bf r}_3})d{\bf r}_3\int\frac{\partial {f_2}({{\bf p}_1},{{\bf p}_2})}{\partial {\bf p}_1}\partial{\bf p}_1\partial{\bf p}_2=0,
\end{multline}
that, when divided by $\int\frac{\partial {f_2}({{\bf p}_1},{{\bf p}_2})}{\partial {\bf p}_1}\partial{\bf p}_1\partial{\bf p}_2\neq0$ gives
\begin{equation}
\label{Ef1}
kT_{eff}\frac{\partial g_{2}({{\bf r}_1},{{\bf r}_2})}{\partial {\bf r}_1}+\frac{\partial \Phi(|{{\bf r}_1}-{{\bf r}_2}|)}{\partial {\bf r}_1}g_{2}({{\bf r}_1},{{\bf r}_2})+\rho\int\frac{\partial \Phi(|{{\bf r}_1}-{{\bf r}_3}|)}{\partial {\bf r}_1}g_{3}({{\bf r}_1},{{\bf r}_2},{{\bf r}_3})d{\bf r}_3=0,
\end{equation}
with
\begin{equation}
\label{Ef2}
kT_{eff}\int\frac{\partial {f_2}({{\bf p}_1},{{\bf p}_2})}{\partial{\bf p}_1}d{\bf p}_1d{\bf p}_2=-\int\frac{{\bf p}_1}{m}{f_2}({{\bf p}_1},{{\bf p}_2})d{{\bf p}_1}d{{\bf p}_2}.
\end{equation}
\end{widetext}
At this point, a new characteristic of the nonequilibrium system in the stationary state can be introduced. Equation (\ref{Ef2}) defines the effective temperature $kT_{eff}$ that, in the general case, differs from the real temperature of the system. It is equal to the temperature of the corresponding equilibrium system with the thermodynamic properties equal to those of the nonequilibrium system under study.

In the particular case when the distribution functions $f_2({{\bf p}_1},{{\bf p}_2})$  and $f_3({{\bf p}_1},{{\bf p}_2},{{\bf p}_3})$ become even, the uncertainty $\frac{0}{0}$ appears in Eq. (\ref{Ef2}). Such a situation can be observed in the case of uniformly distributed sources of irradiation in the liquid. To eliminate the uncertainty, let us take the momentum distribution function as a factor $f_2({\bf{p}_1},{\bf{p}_2})=f_1({\bf{p}_1})f_1({\bf{p}_2})$ and integrate  Eq. (\ref{Ef2}) with respect to $\bf{p_2}$
\begin{equation}
\label{Ef3}
kT_{eff}\frac{\partial {f_1}({{\bf p}_1})}{\partial{\bf p}_1}=-\frac{{\bf p}_1}{m}{f_1}({{\bf p}_1}),
\end{equation}
where $f_1({{\bf p}_1})$ is an even function of ${{\bf p}_1}$. In this case, one may take  $f_1({{\bf p}_1})=\psi(\bf{p}^2)$  and $\frac{\partial {f_1}({{\bf p}_1})}{\partial{\bf p}_1}=2\bf{p}\psi'(p^2)$. Integration of Eq. (\ref{Ef3}) with respect to $\bf{p_1}$ gives
\begin{equation}
\label{Ef4}
kT_{eff}=-\frac{1}{2m}\frac{\int d{{\bf p}_1}{\psi}({{\bf p}_1^{2}})}{\int d{{\bf p}_1}{\psi'}({{\bf p}_1^{2}})}=-\left(2m\int d{{\bf p}_1}{\psi'}({{\bf p}_1^{2}})\right)^{-1}.
\end{equation}
It can be easily seen that when $f_1({{\bf p}_1})$ corresponds to a Maxwellian distribution $\left(f(p)=\frac{1}{\sqrt{2\pi mk_{B}T}}e^{-\frac{p^2}{2mk_{B}T}} \right),$ the effective temperature is equal to the real thermodynamic temperature of the system.

Equation (\ref{Ef4}) is obtained from the second equation of the Bogolyubov chain (Eq. \ref{Bogolyubov1}), defining the pair distribution function. For the majority of systems, this should be sufficient to describe structural changes in the liquid systems, and the result should be the same even when considering higher order distribution functions. As a check,  we have performed the same analysis for the triple distribution function. The third equation of the Bogolyubov chain of equations reads:
\begin{widetext}
\begin{multline}
\label{high1}
\frac{
 \partial{
  F_3\left(
   {\bf{r}}_{1},{\bf{r}}_2,{\bf{r}}_3,{\bf{p}}_1,{\bf{p}}_2,{\bf{p}}_3,t
  \right)
 }
}{
 \partial{t}
}=\left[
 \left(
 \begin{array}{c}
 \frac{\bf{p}_{1}^{2}}{2m}
 +\frac{\bf{p}_{2}^{2}}{2m}
 +\frac{\bf{p}_{3}^{2}}{2m}\\
 +\Phi \left(
  \left|
   {\bf{r}}_1-{\bf{r}}_2
  \right|
 \right)+\Phi \left(
  \left|
   {\bf{r}}_1-{\bf{r}}_3
  \right|
 \right)+\Phi\left(
  \left|
   {\bf{r}}_2-{\bf{r}}_3
  \right|
 \right)
 \end{array}
 \right),
 F_3 \left(
  {\bf{r}}_1,{\bf{r}}_2,{\bf{r}}_3,{\bf{p}}_1,{\bf{p}}_2,{\bf{p}}_3,t
 \right)
\right] \\
+\rho \sum\limits_{i=1}^{n}{
 \int{
  \left[
   \Phi \left(
    \left|
     {\bf{r}}_1-{\bf{r}}_2
    \right|
   \right)+\Phi \left(
    \left|
     {\bf{r}}_1-{\bf{r}}_3
    \right|
   \right)+\Phi \left(
    \left|
     {\bf{r}}_2-{\bf{r}}_3
    \right|
   \right),
   F_4\left(
    {\bf{r}}_1,{\bf{r}}_2,{\bf{r}}_3,{\bf{r}}_4,{\bf{p}}_1,{\bf{p}}_2,{\bf{p}}_3,{\bf{p}}_4,t
   \right)
  \right]
 }
}d{{\bf{r}}_4}d{{\bf{p}}_4},
\end{multline}
\end{widetext}
where $\Phi \left( \left| {{{\bf{r}}}_{1}}-{{{\bf{r}}}_{2}} \right| \right),\Phi \left( \left| {{{\bf{r}}}_{1}}-{{{\bf{r}}}_{3}} \right| \right),\Phi \left( \left| {{{\bf{r}}}_{2}}-{{{\bf{r}}}_{3}} \right| \right)$ are the interaction potentials, and ${{F}_{3}}\left( {{{\bf{r}}}_{1}},{{{\bf{r}}}_{2}},{{{\bf{r}}}_{3}},{{{\bf{p}}}_{1}},{{{\bf{p}}}_{2}},{{{\bf{p}}}_{3}},t \right)$ and ${{F}_{4}}\left( {{{\bf{r}}}_{1}},{{{\bf{r}}}_{2}},{{{\bf{r}}}_{3}},{{{\bf{r}}}_{4}},{{{\bf{p}}}_{1}},{{{\bf{p}}}_{2}},{{{\bf{p}}}_{3}},{{{\bf{p}}}_{4}},t \right)$ are nonequilibrium distribution functions of the third and fourth order respectively.

Opening Poisson brackets in Eq. (\ref{high1}) with the nonequilibrium distribution function of the third order being a factor ${{F}_{3}}\left( {{{\bf{r}}}_{1}},{{{\bf{r}}}_{2}},{{{\bf{r}}}_{3}},{{{\bf{p}}}_{1}},{{{\bf{p}}}_{2}},{{{\bf{p}}}_{3}},t \right)={{g}_{3}}\left( {{{\bf{r}}}_{1}},{{{\bf{r}}}_{2}},{{{\bf{r}}}_{3}} \right){{f}_{3}}\left( {{{\bf{p}}}_{1}},{{{\bf{p}}}_{2}},{{{\bf{p}}}_{3}} \right)$ for the stationary nonequilibrium state ($\frac{\partial {{F}_{3}}\left( {{{\bf{r}}}_{1}},{{{\bf{r}}}_{2}},{{{\bf{r}}}_{3}},{{{\bf{p}}}_{1}},{{{\bf{p}}}_{2}},{{{\bf{p}}}_{3}},t \right)}{\partial t}=0$), one gets
\begin{widetext}
\begin{multline}
\label{high2}
   \left( \frac{\partial \Phi \left( \left| {{{\bf{r}}}_{1}}-{{{\bf{r}}}_{2}} \right| \right)}{\partial {{{\bf{r}}}_{1}}}+\frac{\partial \Phi \left( \left| {{{\bf{r}}}_{1}}-{{{\bf{r}}}_{3}} \right| \right)}{\partial {{{\bf{r}}}_{1}}} \right)\frac{\partial {{f}_{3}}\left( {{{\bf{p}}}_{1}},{{{\bf{p}}}_{2}},{{{\bf{p}}}_{3}} \right)}{\partial {{{\bf{p}}}_{1}}}{{g}_{3}}\left( {{{\bf{r}}}_{1}},{{{\bf{r}}}_{2}},{{{\bf{r}}}_{3}} \right)
  -\frac{{{{\bf{p}}}_{1}}}{m}{{f}_{3}}\left( {{{\bf{p}}}_{1}},{{{\bf{p}}}_{2}},{{{\bf{p}}}_{3}} \right)\frac{\partial {{g}_{3}}\left( {{{\bf{r}}}_{1}},{{{\bf{r}}}_{2}},{{{\bf{r}}}_{3}} \right)}{\partial {{{\bf{r}}}_{1}}} \\
  +\left( \frac{\partial \Phi \left( \left| {{{\bf{r}}}_{1}}-{{{\bf{r}}}_{2}} \right| \right)}{\partial {{{\bf{r}}}_{2}}}+\frac{\partial \Phi \left( \left| {{{\bf{r}}}_{2}}-{{{\bf{r}}}_{3}} \right| \right)}{\partial {{{\bf{r}}}_{2}}} \right)\frac{\partial {{f}_{3}}\left( {{{\bf{p}}}_{1}},{{{\bf{p}}}_{2}},{{{\bf{p}}}_{3}} \right)}{\partial {{{\bf{p}}}_{2}}}{{g}_{3}}\left( {{{\bf{r}}}_{1}},{{{\bf{r}}}_{2}},{{{\bf{r}}}_{3}} \right)
 -\frac{{{{\bf{p}}}_{2}}}{m}{{f}_{3}}\left( {{{\bf{p}}}_{1}},{{{\bf{p}}}_{2}},{{{\bf{p}}}_{3}} \right)\frac{\partial {{g}_{3}}\left( {{{\bf{r}}}_{1}},{{{\bf{r}}}_{2}},{{{\bf{r}}}_{3}} \right)}{\partial {{{\bf{r}}}_{2}}} \\
  +\left( \frac{\partial \Phi \left( \left| {{{\bf{r}}}_{1}}-{{{\bf{r}}}_{3}} \right| \right)}{\partial {{{\bf{r}}}_{3}}}+\frac{\partial \Phi \left( \left| {{{\bf{r}}}_{2}}-{{{\bf{r}}}_{3}} \right| \right)}{\partial {{{\bf{r}}}_{3}}} \right)\frac{\partial {{f}_{3}}\left( {{{\bf{p}}}_{1}},{{{\bf{p}}}_{2}},{{{\bf{p}}}_{3}} \right)}{\partial {{{\bf{p}}}_{3}}}{{g}_{3}}\left( {{{\bf{r}}}_{1}},{{{\bf{r}}}_{2}},{{{\bf{r}}}_{3}} \right)
 -\frac{{{{\bf{p}}}_{3}}}{m}{{f}_{3}}\left( {{{\bf{p}}}_{1}},{{{\bf{p}}}_{2}},{{{\bf{p}}}_{3}} \right)\frac{\partial {{g}_{3}}\left( {{{\bf{r}}}_{1}},{{{\bf{r}}}_{2}},{{{\bf{r}}}_{3}} \right)}{\partial {{{\bf{r}}}_{3}}} \\
  +\rho \int{\left( \begin{array}{l}
   \frac{\partial \Phi \left( \left| {{{\bf{r}}}_{1}}-{{{\bf{r}}}_{4}} \right| \right)}{\partial {{{\bf{r}}}_{1}}}\frac{\partial {{f}_{4}}\left( {{{\bf{p}}}_{1}},{{{\bf{p}}}_{2}},{{{\bf{p}}}_{3}},{{{\bf{p}}}_{4}} \right)}{\partial {{{\bf{p}}}_{1}}}{{g}_{4}}\left( {{{\bf{r}}}_{1}},{{{\bf{r}}}_{2}},{{{\bf{r}}}_{3}},{{{\bf{r}}}_{4}} \right) \\
  +\frac{\partial \Phi \left( \left| {{{\bf{r}}}_{2}}-{{{\bf{r}}}_{4}} \right| \right)}{\partial {{{\bf{r}}}_{2}}}\frac{\partial {{f}_{4}}\left( {{{\bf{p}}}_{1}},{{{\bf{p}}}_{2}},{{{\bf{p}}}_{3}},{{{\bf{p}}}_{4}} \right)}{\partial {{{\bf{p}}}_{2}}}{{g}_{4}}\left( {{{\bf{r}}}_{1}},{{{\bf{r}}}_{2}},{{{\bf{r}}}_{3}},{{{\bf{r}}}_{4}} \right) \\
  +\frac{\partial \Phi \left( \left| {{{\bf{r}}}_{3}}-{{{\bf{r}}}_{4}} \right| \right)}{\partial {{{\bf{r}}}_{3}}}\frac{\partial {{f}_{4}}\left( {{{\bf{p}}}_{1}},{{{\bf{p}}}_{2}},{{{\bf{p}}}_{3}},{{{\bf{p}}}_{4}} \right)}{\partial {{{\bf{p}}}_{3}}}{{g}_{4}}\left( {{{\bf{r}}}_{1}},{{{\bf{r}}}_{2}},{{{\bf{r}}}_{3}},{{{\bf{r}}}_{4}} \right)
\end{array} \right)}d{{{\bf{r}}}_{4}}d{{{\bf{p}}}_{4}}=0.
\end{multline}
Similarly to the case of the pair distribution functions [Eqs. (\ref{Bogolyubov3}) and (\ref{Bogolyubov3a})] from Eq. (\ref{high2}), it is possible to write  three equations for independent coordinates of the form
\begin{multline}
\label{high3}
   \left( \frac{\partial \Phi \left( \left| {{{\bf{r}}}_{1}}-{{{\bf{r}}}_{2}} \right| \right)}{\partial {{{\bf{r}}}_{1}}}+\frac{\partial \Phi \left( \left| {{{\bf{r}}}_{1}}-{{{\bf{r}}}_{3}} \right| \right)}{\partial {{{\bf{r}}}_{1}}} \right)\frac{\partial {{f}_{3}}\left( {{{\bf{p}}}_{1}},{{{\bf{p}}}_{2}},{{{\bf{p}}}_{3}} \right)}{\partial {{{\bf{p}}}_{1}}}{{g}_{3}}\left( {{{\bf{r}}}_{1}},{{{\bf{r}}}_{2}},{{{\bf{r}}}_{3}} \right)
 -\frac{{{{\bf{p}}}_{1}}}{m}{{f}_{3}}\left( {{{\bf{p}}}_{1}},{{{\bf{p}}}_{2}},{{{\bf{p}}}_{3}} \right)\frac{\partial {{g}_{3}}\left( {{{\bf{r}}}_{1}},{{{\bf{r}}}_{2}},{{{\bf{r}}}_{3}} \right)}{\partial {{{\bf{r}}}_{1}}} \\
  +\rho \int{\left( \frac{\partial \Phi \left( \left| {{{\bf{r}}}_{1}}-{{{\bf{r}}}_{4}} \right| \right)}{\partial {{{\bf{r}}}_{1}}}\frac{\partial {{f}_{4}}\left( {{{\bf{p}}}_{1}},{{{\bf{p}}}_{2}},{{{\bf{p}}}_{3}},{{{\bf{p}}}_{4}} \right)}{\partial {{{\bf{p}}}_{1}}}{{g}_{4}}\left( {{{\bf{r}}}_{1}},{{{\bf{r}}}_{2}},{{{\bf{r}}}_{3}},{{{\bf{r}}}_{4}} \right) \right)}d{{{\bf{r}}}_{4}}d{{{\bf{p}}}_{4}}=0.
\end{multline}
Integration of Eq. (\ref{high3}) or of one of the other two equations of that type with respect to ${\bf{p}}_1$, ${\bf{p}}_2$, and ${\bf{p}}_3$, gives the modified equation of the Bogolyubov chain of equations for the nonequilibrium stationary state:
\begin{multline}
\label{high4}
   \left( \frac{\partial \Phi \left( \left| {{{\bf{r}}}_{1}}-{{{\bf{r}}}_{2}} \right| \right)}{\partial {{{\bf{r}}}_{1}}}+\frac{\partial \Phi \left( \left| {{{\bf{r}}}_{1}}-{{{\bf{r}}}_{3}} \right| \right)}{\partial {{{\bf{r}}}_{1}}} \right){{g}_{3}}\left( {{{\bf{r}}}_{1}},{{{\bf{r}}}_{2}},{{{\bf{r}}}_{3}} \right)\int{\frac{\partial {{f}_{3}}\left( {{{\bf{p}}}_{1}},{{{\bf{p}}}_{2}},{{{\bf{p}}}_{3}} \right)}{\partial {{{\bf{p}}}_{1}}}d{{{\bf{p}}}_{1}}d{{{\bf{p}}}_{2}}d{{{\bf{p}}}_{3}}} \\
  -\frac{\partial {{g}_{3}}\left( {{{\bf{r}}}_{1}},{{{\bf{r}}}_{2}},{{{\bf{r}}}_{3}} \right)}{\partial {{{\bf{r}}}_{1}}}\int{\frac{{{{\bf{p}}}_{1}}}{m}{{f}_{3}}\left( {{{\bf{p}}}_{1}},{{{\bf{p}}}_{2}},{{{\bf{p}}}_{3}} \right)}d{{{\bf{p}}}_{1}}d{{{\bf{p}}}_{2}}d{{{\bf{p}}}_{3}} \\
  +\rho \int{\left( \frac{\partial \Phi \left( \left| {{{\bf{r}}}_{1}}-{{{\bf{r}}}_{4}} \right| \right)}{\partial {{{\bf{r}}}_{1}}}{{g}_{4}}\left( {{{\bf{r}}}_{1}},{{{\bf{r}}}_{2}},{{{\bf{r}}}_{3}},{{{\bf{r}}}_{4}} \right)d{{{\bf{r}}}_{4}} \right)\int{\frac{\partial {{f}_{3}}\left( {{{\bf{p}}}_{1}},{{{\bf{p}}}_{2}},{{{\bf{p}}}_{3}} \right)}{\partial {{{\bf{p}}}_{1}}}d{{{\bf{p}}}_{1}}d{{{\bf{p}}}_{2}}d{{{\bf{p}}}_{3}}}}=0,
\end{multline}
\end{widetext}
that when divided by $\int{\frac{\partial {{f}_{3}}\left( {{{\bf{p}}}_{1}},{{{\bf{p}}}_{2}},{{{\bf{p}}}_{3}} \right)}{\partial {{{\bf{p}}}_{1}}}d{{{\bf{p}}}_{1}}d{{{\bf{p}}}_{2}}d{{{\bf{p}}}_{3}}}\neq0$ gives
\begin{multline}
\label{high5}
   k{{T}_{eff}}\frac{\partial {{g}_{3}}\left( {{{\bf{r}}}_{1}},{{{\bf{r}}}_{2}},{{{\bf{r}}}_{3}} \right)}{\partial {{{\bf{r}}}_{1}}}\\
  +\left( \frac{\partial \Phi \left( \left| {{{\bf{r}}}_{1}}-{{{\bf{r}}}_{2}} \right| \right)}{\partial {{{\bf{r}}}_{1}}}+\frac{\partial \Phi \left( \left| {{{\bf{r}}}_{1}}-{{{\bf{r}}}_{3}} \right| \right)}{\partial {{{\bf{r}}}_{1}}} \right){{g}_{3}}\left( {{{\bf{r}}}_{1}},{{{\bf{r}}}_{2}},{{{\bf{r}}}_{3}} \right)+ \\
  +\rho \int{\left( \frac{\partial \Phi \left( \left| {{{\bf{r}}}_{1}}-{{{\bf{r}}}_{4}} \right| \right)}{\partial {{{\bf{r}}}_{1}}}{{g}_{4}}\left( {{{\bf{r}}}_{1}},{{{\bf{r}}}_{2}},{{{\bf{r}}}_{3}},{{{\bf{r}}}_{4}} \right)d{{{\bf{r}}}_{4}} \right)}=0,
\end{multline}
with
\begin{equation}
\label{high6}
-\frac{\int{\frac{{{{\bf{p}}}_{1}}}{m}{{f}_{3}}\left( {{{\bf{p}}}_{1}},{{{\bf{p}}}_{2}},{{{\bf{p}}}_{3}} \right)}d{{{\bf{p}}}_{1}}d{{{\bf{p}}}_{2}}d{{{\bf{p}}}_{3}}}{\int{\frac{\partial {{f}_{3}}\left( {{{\bf{p}}}_{1}},{{{\bf{p}}}_{2}},{{{\bf{p}}}_{3}} \right)}{\partial {{{\bf{p}}}_{1}}}d{{{\bf{p}}}_{1}}d{{{\bf{p}}}_{2}}d{{{\bf{p}}}_{3}}}}=k{{T}_{eff}}.
\end{equation}
The obtained Eq. (\ref{high6}) has the same form as Eq. (\ref{Ef2}), defining the effective temperature in the case of the second order distribution function. In the particular case of the even distribution functions, similarly to the previously described pair distribution function case the triple momentum distribution functions in Eq. (\ref{high6}) can be factorized as $f_3\left({\bf{p}}_1,{\bf{p}}_2,{\bf{p}}_3\right)=f_1({\bf{p}}_1)f_1({\bf{p}}_2)f_1({\bf{p}}_3)$. For this case, by integrating Eq. (\ref{high6}) with respect to ${\bf{p}}_2$, ${\bf{p}}_3$ and ${\bf{p}}_1$ for the effective temperature, one gets the equation
\begin{multline}
\label{high7}
k{{T}_{eff}}=-\frac{\int{\frac{{{{\bf{p}}}_{1}}}{m}{{f}_{3}}\left( {{{\bf{p}}}_{1}},{{{\bf{p}}}_{2}},{{{\bf{p}}}_{3}} \right)}d{{{\bf{p}}}_{1}}d{{{\bf{p}}}_{2}}d{{{\bf{p}}}_{3}}}{\int{\frac{\partial {{f}_{3}}\left( {{{\bf{p}}}_{1}},{{{\bf{p}}}_{2}},{{{\bf{p}}}_{3}} \right)}{\partial {{{\bf{p}}}_{1}}}d{{{\bf{p}}}_{1}}d{{{\bf{p}}}_{2}}d{{{\bf{p}}}_{3}}}} \\
=-\frac{\int{\frac{{{{\bf{p}}}_{1}}}{m}{{f}_{1}}\left( {{{\bf{p}}}_{1}} \right)}d{{{\bf{p}}}_{1}}}{\int{\frac{\partial {{f}_{1}}\left( {{{\bf{p}}}_{1}} \right)}{\partial {{{\bf{p}}}_{1}}}d{{{\bf{p}}}_{1}}}}
\end{multline}
which is of the same form as that for the case of binary distribution function considered (Eq. (\ref{Ef3})). Therefore, for studying thermophysical properties of the liquid systems under irradiation it is sufficient to consider only the second equation from the Bogolyubov chain. Unfortunately, for the case of strongly interacting subsystems when the interaction is dependent on the particles velocity the situation becomes more complicated and requires the inclusion of  distribution functions of higher orders.

\subsection{General description}
An analysis for the two-component liquid system under irradiation can be found, e.g., in Ref. \cite{Vlasenko2014}. The results of the analysis show that in a two-component system, three different effective temperatures must exist:

\begin{align}
\label{Efftwo}
&-\frac{\int{\frac{\bf{p}_{1}^{l_i}}{{{m}_{l_i}}}{{f}_{2}}\left( \bf{p}_{1}^{l_i},\bf{p}_{2}^{l_i} \right)}d\bf{p}_{1}^{l_i}d\bf{p}_{2}^{l_i}}{\int{\frac{\partial {{f}_{2}}\left( \bf{p}_{1}^{l_i},\bf{p}_{2}^{l_i} \right)}{\partial \bf{p}_{1}^{l_i}}d\bf{p}_{1}^{l_i}d\bf{p}_{2}^{l_i}}}=kT_{eff}^{l_il_i}, \notag \\
&-\frac{\int{\frac{\bf{p}_{1}^{l_j}}{{{m}_{l_j}}}{{f}_{2}}\left( \bf{p}_{1}^{l_j},\bf{p}_{2}^{l_j} \right)}d\bf{p}_{1}^{l_j}d\bf{p}_{2}^{l_j}}{\int{\frac{\partial {{f}_{2}}\left( \bf{p}_{1}^{l_j},\bf{p}_{2}^{l_j} \right)}{\partial \bf{p}_{1}^{l_j}}d\bf{p}_{1}^{l_j}d\bf{p}_{2}^{l_j}}}=kT_{eff}^{l_jl_j},\notag \\
&-\frac{\int{\frac{\bf{p}_{1}^{l_i}}{{{m}_{l_i}}}{{f}_{2}}\left( \bf{p}_{1}^{l_i},\bf{p}_{2}^{l_j} \right)}d\bf{p}_{1}^{l_i}d\bf{p}_{2}^{l_j}}{\int{\frac{\partial {{f}_{2}}\left( \bf{p}_{1}^{l_i},\bf{p}_{2}^{l_j} \right)}{\partial \bf{p}_{1}^{l_i}}d\bf{p}_{1}^{l_i}d\bf{p}_{2}^{l_j}}}=kT_{eff}^{l_il_j},
\end{align}
where $l_i$ and $l_j$ denote the types of particles and $kT_{eff}^{l_il_i}$, $kT_{eff}^{l_jl_j}$ and $kT_{eff}^{l_il_j}=kT_{eff}^{l_jl_i}$ are the effective temperatures of the $l_il_i$, $l_jl_j,$ and $l_il_j$ subsystems, respectively. Such qualitative results have much in common with the situation well known in the statistical theory of relaxation processes of the systems consisting of subsystems with weak interaction \cite{Zubarev1997}. In such systems, it is quite common to have different temperatures of the subsystems. It can be easily seen from Eqs. (\ref{Efftwo}) that for the case of an even distribution function, $kT_{eff}^{l_il_j}$ vanishes and the system is characterized by two effective temperatures.

The suggested approach can be naturally extended to multicomponent systems. In the case of the
$M-$component system, one can write a set of Bogolyubov chains of equations for the $s$-th order distribution functions as:
\begin{widetext}
\begin{multline}
\label{Bogolyubovmult}
   \frac{\partial {{F}_{s}^{l_1...l_M}}}{\partial t}=\left[
   \left(
    \begin{array}{l}
     \frac{\left({p}_{1}^{l_1}\right)^2}{2m^{l_1}}+...+\frac{\left({p}_{s}^{l_M}\right)^2}{2m^{l_M}}\\
     +\Phi^{l_1l_2}\left(\left|{\bf{r}}_{1}^{l_1}-{\bf{r}}_{2}^{l_2}\right| \right)+...+\Phi^{l_{M-1}l_M}\left(\left|{\bf{r}}_{s-1}^{l_{M-1}}-{\bf{r}}_{s}^{l_M}\right| \right)
    \end{array}
   \right),{F}_{s}^{l_1...l_M}\right] \\
   +\sum\limits_{k=1}^{M}{
      \rho_k \int{
       \left(
        \frac{\partial\Phi^{l_1k}\left(\left|{\bf{r}}_{1}^{l_1}-{\bf{r}}_{s+1}^{k}\right| \right)}{\partial{\bf{r}}_{1}^{l_1}}\frac{\partial{{F}_{s+1}^{l_1...l_M,k}}}{\partial{\bf{p}}_{1}^{l_1}}
        +...+
        \frac{\partial\Phi^{l_Mk}\left(\left|{\bf{r}}_{s}^{l_M}-{\bf{r}}_{s+1}^{k}\right| \right)}{\partial{\bf{r}}_{s}^{l_M}}\frac{\partial{{F}_{s+1}^{l_1...l_M,k}}}{\partial{\bf{p}}_{s}^{l_M}}
       \right)d{\bf{r}}_{s+1}^{k}d{\bf{p}}_{s+1}^{k}}
      },
\end{multline}
with $l_1...l_M$ denoting all possible types of components, ${{F}_{s}^{l_1...l_M}}={{F}_{s}^{l_1...l_M}}\left({\bf{r}}_{1}^{l_1},...,{\bf{r}}_{s}^{l_M},{\bf{p}}_{1}^{l_1},...,{\bf{p}}_{s}^{l_M},t\right)$ and ${{F}_{s+1}^{l_1...l_M,k}}={{F}_{s}^{l_1...l_M}}\left({\bf{r}}_{1}^{l_1},...,{\bf{r}}_{s}^{l_M},{\bf{r}}_{s+1}^{k},{\bf{p}}_{1}^{l_1},...,{\bf{p}}_{s}^{l_M},{\bf{p}}_{s+1}^{k},t\right)$. If one cuts the Bogolyubov chain at the second equation, then for the stationary nonequilibrium state it is possible to obtain the set of the equations for the binary distribution functions in the form:
\begin{multline}
\label{Bogolyubovmult1}
   \frac{{\bf{p}}_{1}^{l_i}}{m^{l_i}}\frac{\partial {{F}_{2}^{l_il_j}}}{\partial {\bf{r}}_{1}^{l_i}}+
  \frac{{\bf{p}}_{2}^{l_j}}{m^{l_j}}\frac{\partial {{F}_{2}^{l_il_j}}}{\partial {\bf{r}}_{2}^{l_j}}-
  \frac{\partial{{F}_{2}^{l_il_j}}}{\partial{\bf{p}}_{1}^{l_i}}\frac{\partial\Phi^{l_il_j}\left(\left|{\bf{r}}_{1}^{l_i}-{\bf{r}}_{2}^{l_j}\right| \right)}{\partial{\bf{r}}_{1}^{l_i}}-
  \frac{\partial{{F}_{2}^{l_il_j}}}{\partial{\bf{p}}_{2}^{l_j}}\frac{\partial\Phi^{l_il_j}\left(\left|{\bf{r}}_{1}^{l_i}-{\bf{r}}_{2}^{l_j}\right| \right)}{\partial{\bf{r}}_{2}^{l_j}} \\
   =\sum\limits_{k=1}^{M}{
      \rho_k \int{
       \left(
        \frac{\partial\Phi^{l_ik}\left(\left|{\bf{r}}_{1}^{l_i}-{\bf{r}}_{3}^{k}\right| \right)}{\partial{\bf{r}}_{1}^{l_i}}\frac{\partial{{F}_{3}^{l_il_jk}}}{\partial{\bf{p}}_{1}^{l_i}}+
        \frac{\partial\Phi^{l_jk}\left(\left|{\bf{r}}_{2}^{l_j}-{\bf{r}}_{3}^{k}\right| \right)}{\partial{\bf{r}}_{2}^{l_j}}\frac{\partial{{F}_{3}^{l_il_jk}}}{\partial{\bf{p}}_{2}^{l_j}}
       \right)d{\bf{r}}_{3}^{k}d{\bf{p}}_{3}^{k}
      }
     }
\end{multline}
with $l_i,l_j$ defining any possible type of the components.
\end{widetext}
All of these equations have forms similar to the case of a single-component system (Eq. (\ref{Bogolyubov2})). Skipping the math that is similar to the single-component case, it is possible to calculate the new characteristics of the system that are effective temperatures:
\begin{align}
\label{Effmult}
  & kT_{eff}^{l_il_j}=-\frac{\int{\frac{{\bf{p}}_{1}^{l_i}}{{{m}_{l_i}}}{{f}_{2}}\left( {\bf{p}}_{1}^{l_1},{\bf{p}}_{2}^{l_j} \right)}d{\bf{p}}_{1}^{l_i}d{\bf{p}}_{2}^{l_j}}{\int{\frac{\partial {{f}_{2}}\left( {\bf{p}}_{1}^{l_i},{\bf{p}}_{2}^{l_j} \right)}{\partial {\bf{p}}_{1}^{l_i}}d{\bf{p}}_{1}^{l_i}d{\bf{p}}_{2}^{l_j}}} \notag \\
  & l_i,l_j=1...M
\end{align}
It can be seen that, similarly to the two-component system, $kT_{eff}^{l_il_j}=kT_{eff}^{l_jl_i}$. The overall number of the effective temperatures in the general case is $M+\frac{M!}{\left( M-2 \right)!}$ with $M\geq2$ being the number of components.

\section{Results and discussion.}
 One can conclude from Eqs. (\ref{Ef1})-(\ref{Ef4}) that in order to have a detailed description of a nonequilibrium liquid system under irradiation in the stationary state, it is necessary to know the distorted velocity distribution function of the system. Our approach provides a link between the structural and the thermodynamic properties of the nonequilibrium system. Within the introduced model, the  knowledge of the coefficients $A$ and $\phi$ of the modified Maxwell distribution function allows one to calculate the thermodynamic properties of the nonequilibrium stationary system under irradiation. Such properties should be the same with the properties of the corresponding equilibrium system with $T=T_{eff}$.

In the model presented, the changes in the thermodynamic parameters of the irradiated system depend on the effective temperature, which in turn depends on the parameters $A$ and $\phi$ of the modified Maxwell distribution function. To quantify the changes of properties in the system caused by the changes in the momentum distribution function, we have calculated effective temperatures for a number of model systems. To do this we use the classical Maxwell distribution function multiplied by the modified orthogonal Hermite polynomial of the fourth order,
\begin{equation}
\label{Mod1}
F(p)=\frac{1}{A^{\frac{1}{2}}\pi^{\frac{1}{2}}}\exp^{-\frac{p^2}{A}}\left(\frac{q}{A^2}p^4+\frac{d}{2A}p^2+1\right),
\end{equation}
where $A=2mk_{B}T$ and $q$ and $d$ are free parameters. Such a choice seems to be reasonable as the orthogonal Hermite polynomial are the well studied weighted orthogonal functions and the weighting coefficient can be taken in the form of a Maxwell distribution exponent. This makes the model function easy to treat analytically providing physically meaningful results. In this case, from Eqs. (\ref{Ef4}) and (\ref{Mod1}) the effective temperature can be calculated as
\begin{equation}
\label{Mod2}
kT_{eff}=-\frac{8k_{B}T}{q+2d-4}.
\end{equation}
Varying the parameters $q$ and $d$ ($q+2d-4\neq0$), one can get a number of modified velocity distribution functions. The results for some sets of parameters are shown in Table \ref{Temp1}.

\begin{figure}[h]
\center{\includegraphics[scale=1.0]{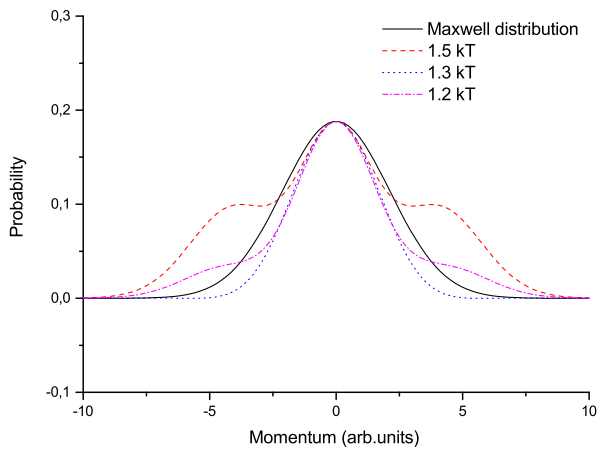}}
\caption{Model momentum distribution functions. The corresponding sets of parameters are given in Table \ref{Temp1}.}
\label{FigTemp}
\end{figure}

\begin{table}[htp]
\caption{Free parameters $q$ and $d$ with correspondent effective temperatures}
\label{Temp1}
 \begin{ruledtabular}
 \begin{tabular}{ccc}
 \textrm{q} & \textrm{d}& \textrm{$kT_{eff}$}\\*[3pt]
 1 & $-\frac{7}{6}$ & 1.5$k_BT$ \\
 $\frac{1}{13}$ & $-\frac{29}{26}$ & 1.3$k_BT$ \\
 $\frac{1}{2}$ & $-\frac{19}{12}$ & 1.2$k_BT$ \\
 0 & 0 & $k_BT$ \\
 \end{tabular}
\end{ruledtabular}
\end{table}
It can be seen from Fig. \ref{FigTemp} and Table \ref{Temp1} that even small variations in the momentum distribution function lead to noticeable changes in the effective temperature and, hence, in the thermodynamic properties of the system.

\begin{figure}[h]
\center{\includegraphics[scale=0.5]{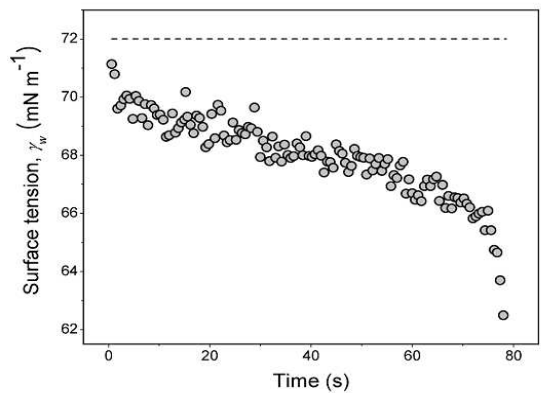}}
\caption{Surface tension coefficient dependence on the irradiation time. X-Ray irradiation with the absorbed dose rate 1000 $Gy{\cdot}s^{-1}$. Irradiation time $\tau=80 s$. \cite{Weon2008}}
\label{FigExp1}
\end{figure}

We have also shown that, for $q=d=0,$ the momentum distribution function is a Maxwell function and the effective temperature is equal to the real thermodynamic temperature of the system.
As a qualitative proof of the suggested approach, we compare the results with the existing experiments studying changes in the thermodynamic parameters of the liquid systems under irradiation.
In a number of experimental works devoted to the studies of the irradiation influence on liquids, changes in surface tension coefficient are observed for high intensity irradiation \cite{Martino2006, Zenkiewicz2007, Byung2008, Weon2008}. At the same time, the measured temperature of the systems stays stable with variations ${\delta}T<1 K$ \cite{Weon2008}. The experimental data are presented in Fig. \ref{FigExp1}.

Explanations of the phenomenon that exist at present \cite{Martino2006, Zenkiewicz2007, Byung2008, Weon2008} give an explanation of the changes in the surface tension coefficient under irradiation. At the same time, they do not provide any general physical picture of the processes responsible for the changes of the thermodynamic characteristics of liquids under irradiation.
Therefore, it might be interesting to compare our model with the existing experimental results.
The surface tension coefficient depends on the permanent structure of the liquid, and that dependence is given by the well-known Fowler equation \cite{Rowlinson1982}
\begin{equation}
\label{Surf1}
\sigma=\frac{\pi}{8}\rho_{l}^{2}\int_{0}^{\infty}d{\bf{r}}_{12}{\bf{r}}_{12}^{4}u'({\bf{r}}_{12})g({\bf{r}}_{12})
\end{equation}
Our model suggests a possible link between the structural and thermodynamic properties of the liquid systems under irradiation.  To compare the results, we calculate the effective temperature that can explain the observed changes in the surface tension coefficients.

\begin{figure}[h]
\center{\includegraphics[scale=0.9]{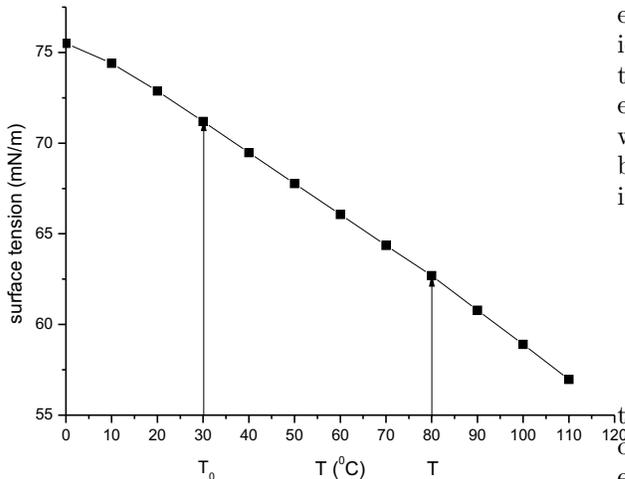}}
\caption{Temperature dependence of the surface tension coefficient \cite{Vargaftik1983}.}
\label{FigRef1}
\end{figure}

The temperature dependence of the surface tension coefficient of water is well known and can be found in the reference data tables \cite{Vargaftik1983} (Fig. \ref{FigRef1}).

To calculate the effective temperature of the water in the experiment \cite{Martino2006, Zenkiewicz2007, Byung2008, Weon2008} it is possible to use the equality
\begin{equation}
\label{Comp}
kT_{eff}=k_{B}T_{0}(1+\frac{\Delta T}{T_{0}}),
\end{equation}
where $\Delta{T}=T-T_{0}$, $Ò_{0}$ is the temperature of the system before irradiation and $Ò$ is the temperature that corresponds to the measured surface tension after the irradiation according to reference data (Fig. \ref{FigRef1}). In the experiment, the surface tension coefficient before irradiation is $\sigma_{0}=71.2 \frac{mN}{m},$ temperature of the system being $T_{0}=303K$, while after irradiation it appears to be $\sigma_i=62.7 \frac{mN}{m}$. According to the reference data (Fig. \ref{FigRef1}), this value $\sigma_i$ corresponds to the temperature $T=353 K$. Then from Eq. (\ref{Comp}) one can easily calculate the effective temperature:
\begin{equation}
\label{Comp2}
kT_{eff}\approx 1.2 k_{B}T
\end{equation}
Such a value of the effective temperature seems to be reasonable, corresponding to small changes in the momentum distribution function as it can be seen from Table \ref{Temp1} and Fig. \ref{FigTemp}.
From the above analysis it can be seen that the existing experimental works qualitatively confirm our approach. Similar results may be obtained from the analysis of the  electroconductivity, linked with the instant structure in the two-component liquid systems \cite{Vlasenko2015}. At the same time, for a detailed quantitative confirmation of the suggested approach, it is necessary to have some precise data on momentum distribution function or the structural changes in the liquid systems under irradiation.

Even though the developed approach is suggested for ordinary liquids, it seems to be possible to use it also for studying radiation influence on dense gases and supercritical state. In that case, further analysis of restricting to the second order distribution functions as well as the possibility to factorize them should be done. Such a statement can be justified by the fact that the method of Bogolyubov chains is applicable for such systems \cite{Green1980} and this suggested approach has no restrictions stemming from the properties of the ordinary liquids. It may be used also for quantifying radiation damage effects in solids that are intensively studied nowadays \cite{Trachenko2006}; however, to check its applicability for that case, further studies are needed. At the same time, it seems that studying the deviations of the solid structure from the reference equilibrium state is a more straightforward and easier way. Therefore, our approach, in its present state, can be used for studying radiation influence on liquid systems in the stationary nonequilibrium state.

\section{Conclusions.}
In this paper, a method to calculate the structural and thermodynamical changes in a liquid systems under irradiation, based on the fundamental Bogolyubov chain of equations, is suggested. Our analysis shows that, even the inclusion of the second equation of the chain alone, provides complete description of the thermophysical and structural changes in the nonequilibrium stationary system in the case when the interaction is independent on the particles' velocities. We suggest that the main mechanism responsible for changes of the system parameters is contained in the modification of the coefficients of the momentum distribution function due to momentum exchange between active particles and those of the liquid system. Our approach gives the possibility to derive equations that relate the modified momentum distribution function with the distorted pair distribution function of the nonequilibrium liquid system in the stationary state.

The proposed approach allows one to introduce a new characteristic of the nonequilibrium liquid system in the stationary state under irradiation, namely an effective temperature. In the general case, that temperature differs from the real measured temperature and corresponds to the temperature of the equilibrium system with the same thermophysical characteristics. It allows us to link the structural changes and the changes in the momentum distribution function with the thermophysical properties of the stationary nonequilibrium system. It allows us also to recover all the thermophysical properties of such a system once the effective temperature is known. In the case of a multicomponent system, several effective temperatures characteristic for the subsystems appear.

A qualitative comparison of the suggested approach with the existing experimental data on thermophysical properties of irradiated liquids suggests that our model is able to explain the observed changes in the surface tension coefficient.

\bibliography{IrradLiquids}

\end{document}